# Quantum transport simulations for the thermoelectric power factor in 2D nanocomposites

Samuel Foster[1][1], Mischa Thesberg[2], and Neophytos Neophytou[1]

[1]*School of Engineering, University of Warwick, Coventry, CV4 7AL, UK*
[2]*Institute for Microelectronics, Technical University of Vienna, Austria*

**Abstract**

Some of the most promising candidates for next-generation thermoelectrics are nanocomposites due to their low thermal conductivities that result from phonon scattering on the boundaries of the various material phases. However, in order to maximize the figure of merit *ZT*, it is important to understand the impact of such features on the thermoelectric power factor. In this work we consider the effect that nanoinclusions and voids have on the electronic and thermoelectric coefficients of two-dimensional geometries using the fully quantum mechanical Non-Equilibrium Green's Function method. This method combines in a unified approach the details of geometry, electron-phonon interactions, quantisation, tunnelling, and the ballistic to diffusive nature of transport. We show that as long as the barrier height is low nanoinclusions can have a positive impact on the Seebeck coefficient and the power factor is not severely impacted by a reduction in conductance. The power factor is also shown to be approximately independent of nanoinclusion and void density in the ballistic case. On the other hand, in the presence of phonon scattering voids degrade the power factor and their influence increases with density.

* Corresponding author.
*E-mail address:* S.Foster@warwick.ac.uk

## 1. Introduction

Thermoelectric materials convert directly between heat differences and potential differences. The performance of such materials is quantified by the dimensionless figure of merit: $ZT = \sigma S^2 T/(\kappa_e + \kappa_l)$ where $\sigma$ is the electrical conductivity, $S$ is the Seebeck coefficient, $T$ is the temperature, $\kappa_e$ is the electron thermal conductivity, $\kappa_l$ is the lattice thermal conductivity, and $\sigma S^2$ is known as the power factor ($PF$).

Traditionally, $ZT$ has been approximately 1, although recently various materials have demonstrated $ZT$ above 1, primarily by the reduction of the thermal conductivity [1]. There are many methods that exist to reduce $\kappa$ beyond bulk values, and one of the most common of these has been the use of nanoinclusions [2,3,4,5,6,7] as well as nanoporous materials, since these cause scattering of short wavelength phonons. While this impact on thermal conductivity is well studied [8,9,10,11], it is not so clear from previous results what impact there is on the power factor in such geometries. Results vary significantly from only small influence [12,13,14,15], to claims of large potential improvements [16,17,18,19]. In our previous work we have studied the impact of nanoinclusions of finite barrier height [20]; we now extend that work and compare it to the influence of voids which can be considered as infinite potential barriers.

In this work we use the fully quantum mechanical Non-Equilibrium Green's Function (NEGF) simulation method to calculate the electronic and thermoelectric transport properties of 2D geometries embedded with nanoinclusions and voids. We show that nanoinclusions can have a positive impact on the Seebeck coefficient, and that consequently the power factor is not severely degraded by reductions in the conductance. In the ballistic regime, we show that the power factor is independent of nanoinclusion/void density, while in the phonon scattering case increasing density has a detrimental effect on the power factor.

Our simulation method is outlined in Section 2. In Section 3 we present and discuss our results before drawing our conclusions in Section 4.

## 2. Approach

To compute the electronic transport we have developed a 2D quantum transport simulator based on the Non-Equilibrium Green's Function (NEGF) formalism including electron-acoustic phonon scattering in the self-consistent Born approximation [21,22]. The system is treated as a 2D channel within the effective mass approximation, using a uniform $m^* = m_0$ throughout the channel, where $m_0$ is the rest mass of the electron. The nanoinclusions are modelled as potential barriers of cylindrical shape within the matrix material as shown in the schematic of Fig. 1, and the voids are modelled as "infinite" potential barriers where we set the potential to 10 eV. The NEGF theory is described adequately in various places in the literature [21,22,23] so we do not include it here. Most work on NEGF in the literature is applied to 1D systems due to computational limitations, however in this work we expand the formalism to 2D systems of widths $W = 30$ nm and lengths $L = 60$ nm. The Recursive Green's Function (RGF) formalism is used to calculate the relevant elements of the Green's function, and the Sancho-Rubio algorithm to compute the self-energies of the contacts [24].

We model the effect of electron scattering with acoustic phonons by including a self-energy on the diagonal elements of the Hamiltonian. The convergence criteria for the ensuing self-consistent calculation is chosen to be current conservation, i.e. we consider convergence is achieved when the current is conserved along the length of the channel to within 1%. The strength of the electron-phonon coupling is given by $D_0$, which we consider uniform across the entire channel.

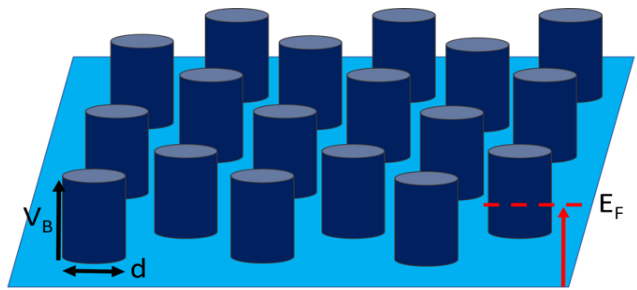

Fig. 1. A schematic of a typical geometry we consider. $V_B$ is the barrier height, $d$ the nanoinclusion diameter, and $E_F$ the Fermi level.

We assume room temperature $T = 300$ K throughout the paper. The value of $D_0$ is chosen such that the conductance of an $L = 15$ nm long pristine channel is found to be 50% of the ballistic value. This effectively amounts to fixing a mean-free-path of 15 nm for the system [25]; a value that is comparable to common semiconductors such as silicon [26,27,28]. Thus, with such a mean-free-path, the $L = 60$ nm channel length we consider is large enough to result in diffusive transport in the material we simulate. The conduction band is set at $E_C = 0.00$ eV and the Fermi level is placed at $E_F = 0.05$ eV.

## 3. Results

Once the calibration of our channel is completed we proceed to consider geometries which include circular nanoinclusions of different barrier heights, $V_B$, voids, different NI/void densities, and different NI/void diameters. The channel width was kept at $W = 30$ nm, and the length at $L = 60$ nm in all cases.

We first consider the ballistic (coherent) scattering regime. The thermoelectric coefficients $G$, $S$ and $PF$, are shown in Fig. 2 versus barrier height $V_B$ for four simulated geometries as shown in the insets of Fig. 2c. The four simulated geometries are: i) a 2×4 array (green lines), ii) a 4×4 array (black lines), iii) a 6×4 array (blue lines), and iv) an 8×4 array (red lines), and the Fermi level is placed at $E_F = 0.05$ eV (dashed-red line in Fig. 2c). Figure 2a shows that, as expected, $G$ decreases both with increasing $V_B$, and with increasing NI/void density. Increasing NI/void density leads to an increase in the Seebeck (shown in Fig. 2b) although in the void case the situation becomes more complicated as coherent resonance effects come into play. The result of the improvement in $S$ is that the power factor (shown in Fig. 2c) increases from the pristine channel value at a barrier height $V_B = 0.05$ eV, before falling again.

The second investigation we perform is on the influences of: i) the nanoinclusion barrier height, $V_B$ (including voids – with effective infinite barrier height) and ii) the density of nanoinclusion/voids, on the thermoelectric coefficients in the acoustic phonon scattering regime. Fig. 3 shows the thermoelectric coefficients conductance $G$, Seebeck coefficient $S$, and power factor $GS^2$ versus $V_B$ for four different geometries of increasing density. The four simulated geometries are again: i) a 2×4 array (green lines), ii) a 4×4 array (black lines), iii) a 6×4 array (blue lines), and iv) an 8×4 array (red lines). The Fermi level is placed at $E_F = 0.05$ eV (dashed-red line in Fig. 3c). From Fig. 3a we can see that, as before, as the barrier height is increased $G$ falls, while $G$ also falls as the density of the nanoinclusions/voids is increased. $S$ shows an initial increase as low energy carriers are filtered out before falling and saturating at an in-between value. Since the Seebeck coefficient $S \propto\, <E - E_F>$ (i.e. $S$ is proportional to the average energy of the current flow) this saturation appears to suggest that after $V_B \sim 2k_BT$ above $E_F$, electrons of all energies contributing to the current are affected relatively equally and we show this later. The continued decrease in $G$ however (comparing $V_B = 0.1$ eV to the voids in Fig. 3a) indicates that there is a further reduction in flow, but that this occurs relatively evenly across the energy range. Fig. 3c shows the results of these effects on the power factor. The initial introduction of a barrier has a reasonably limited effect on the power factor (only a 15% reduction even at the largest NI density) while a further increase of $2k_BT$ in the barrier height produces a more significant reduction (a further 26% at the largest NI density). Interestingly this fall is larger than that from $V_B \sim 0.1$ eV to voids (a further 17% at the largest NI density) reflecting the fact that the majority of the electron flow occurs within $2k_BT$ of $E_F$. Due to the detrimental impact of the NI/voids on $G$, the $PF$ also degrades as the NI/void density is increased, and there is no increase at $V_B = 0.05$ eV as was seen in the ballistic case.

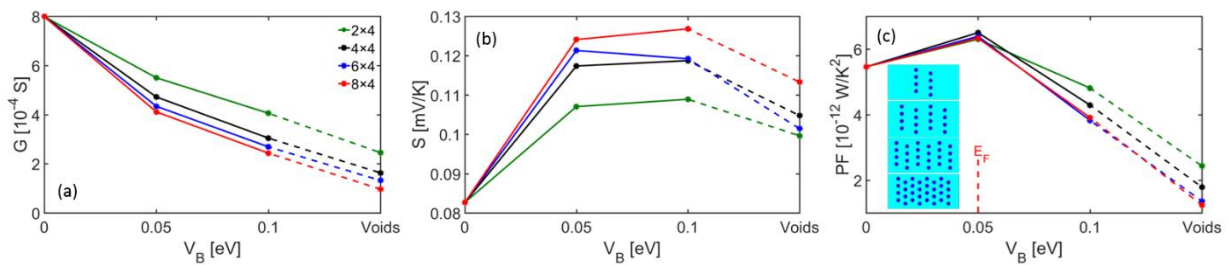

Fig. 2. The thermoelectric coefficients of an $L = 60$ nm channel with $E_F = 0.05$ eV (dashed-red line) and ballistic transport conditions versus nanoinclusion barrier height, $V_B$. (a) The conductance. (b) The Seebeck coefficient. (c) The power factor defined as $GS^2$. Hexagonal arrays of four different nanoinclusion and void densities are considered as shown in the inset of (c): 2×4 array (green lines), 4×4 array (black lines), 6×4 array (blue lines), and 8×4 array (red lines).

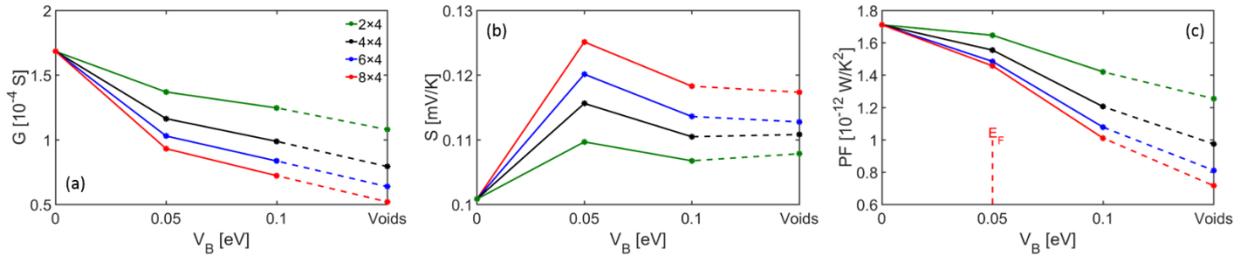

Fig. 3. The thermoelectric coefficients of an $L = 60$ nm channel with $E_F = 0.05$ eV (dashed-red line) and acoustic phonon scattering transport conditions versus nanoinclusion barrier height, $V_B$. (a) The conductance. (b) The Seebeck coefficient. (c) The power factor defined as $GS^2$. Hexagonal arrays of four different nanoinclusion and void densities are considered as shown in the inset of Fig. 2c: 2×4 array (green lines), 4×4 array (black lines), 6×4 array (blue lines), and 8×4 array (red lines).

The next investigation we perform is to illustrate the effects that density and void diameter have on the thermoelectric coefficients. Fig. 4 shows the thermoelectric coefficients $G$, $S$ and $PF$, versus void density for two void diameters: i) $d = 3$ nm (red lines), and ii) $d = 1.5$ nm (black lines). An example geometry for each void diameter is shown in the inset of Fig. 4c. The Fermi level is placed at $E_F = 0.05$ eV, and acoustic phonon scattering is included. As expected, an increase in the void density reduces $G$ and increases $S$. At higher densities resonances and interference effects have an additional detrimental impact on $G$ for the small diameter (since the average distance between the voids becomes smaller than the mean-free-path) and produce an equivalent increase in $S$. The overall effect on the $PF$ is a reduction as expected from Fig. 3. What is important to note, however, is that this reduction is independent of the void diameter, even at higher densities where quantum effects become important.

Finally, to better understand the electronic transport through the geometries we have considered we show in Fig. 5 the transmission and the current as they vary in energy. In Fig. 5a we show the transmission for four different scattering cases: i) the pristine channel in the coherent ballistic regime (blue 'staircase' line), ii) the pristine channel with acoustic phonon scattering (red line), iii) a channel with an 8×4 array of $d = 3$ nm voids in the coherent ballistic regime (light blue line), iv) a channel with an 8×4 array of $d = 3$ nm voids and acoustic phonon scattering (light red line). The ballistic transmission of the pristine channel shows the expected staircase shape, with an increment every time a new subband is reached in energy. When voids are inserted into the geometry the transmission is reduced significantly as well as showing resonance features. Those resonances are smoothened out when phonon scattering is included, and the transmission is reduced even more when voids are added in addition to phonon scattering.

In Fig. 5b we plot the energy-weighted current in the transport direction versus energy for two acoustic phonon scattering different cases: i) a channel with an 8×4 array of $d = 3$ nm nanoinclusions with barrier height $V_B = 0.1$ eV (black line), ii) a channel with an 8×4 array of $d = 3$ nm voids (blue line). It can be seen that changing from nanoinclusions of barrier height $V_B = 0.1$ eV to voids affects electrons of all energies similarly (including those with energies below $V_B = 0.1$ eV). This explains the lack of any change in $S$ as we go from nanoinclusions of barrier height $V_B = 0.1$ eV to voids as shown in Fig. 3b. Because all electron energies are affected to a similar degree, the average energy of the current flow does not change and hence, since $S \propto <E - E_F>$, neither does $S$.

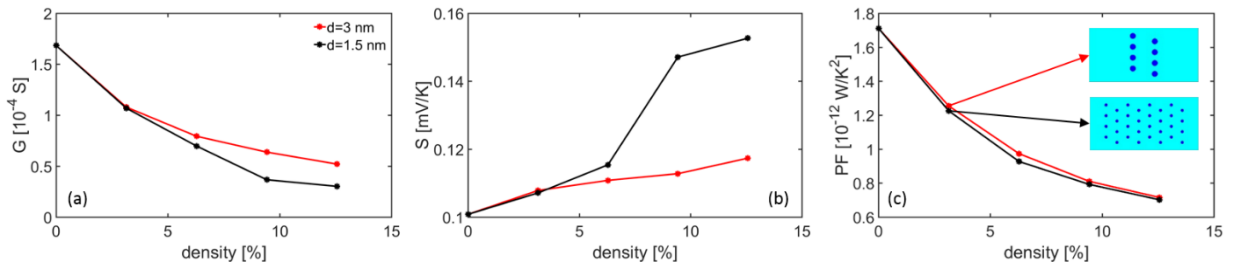

Fig. 4. The thermoelectric coefficients of an $L = 60$ nm channel with $E_F = 0.05$ eV and acoustic phonon scattering transport conditions versus void density. (a) The conductance. (b) The Seebeck coefficient. (c) The power factor defined as $GS^2$. Two different diameters of voids are considered: i) $d = 3$ nm (red lines), ii) $d = 1.5$ nm (black lines). An example geometry for each void diameter is shown in the inset of (c).

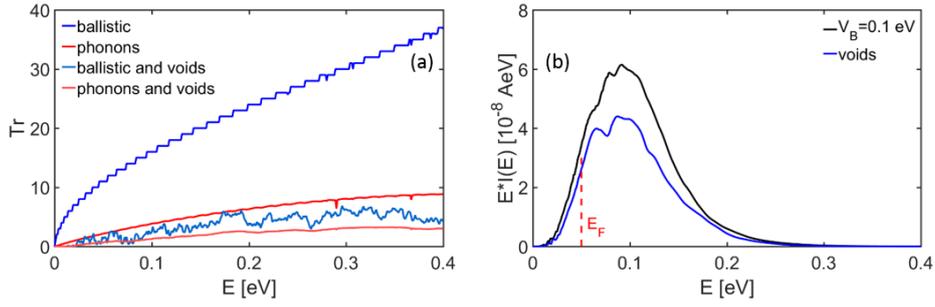

Fig. 5. (a) The transmission versus energy for an $L = 60$ nm channel in four different cases: i) a pristine channel under ballistic conditions (blue line), ii) a pristine channel under acoustic phonon scattering conditions (red line), iii) a channel with an 8×4 hexagonal array of voids under ballistic conditions (light-blue line), iv) a channel with an 8×4 hexagonal array of voids under acoustic phonon scattering conditions (light-red line). (b) The energy-weighted current flow in the transport direction versus energy under acoustic phonon scattering conditions for two cases: i) a channel with an 8×4 hexagonal array of nanoinclusions of barrier height $V_B = 0.1$ eV (black line), ii) a channel with an 8×4 hexagonal array of voids (blue line).

At this point, we would like to comment on the possible consequence of our results for the figure of merit $ZT$. While we have not performed thermal conductivity calculations on the structures we consider (this will require elaborate Molecular Dynamics (MD) or Monte Carlo (MC) simulations), we now qualitatively combine our power factor results with thermal conductivity results found in the literature. In Ref. [29], Dunham *et* al. claim from experiments and MC simulations that small void diameters (~4 nm) in Si channels, similar to those we consider, can result in thermal conductivity reductions of an order of magnitude compared to the bulk. In another work, Lee *et* al. showed from MD simulations that small diameter voids can produce reductions from the bulk value of Si by up to two orders of magnitude [30]. Likewise, MD simulations of nanoporous SiGe have shown thermal conductivity reductions of over an order of magnitude [31]. If we combine this with the halving of the power factor shown in Fig. 4c, we can extract a rough estimate of at least 5x increase in $ZT$. In the case of nanoinclusions, one would expect that the thermal conductivity reduction is not as strong compared to the case of nanovoids, however, in a number of examples where nanoinclusions are formed within matrix materials it is still found that thermal conductivity can be reduced by an order of magnitude [32, 33]. Since in this case we see little change in the power factor (see Fig. 3c), we would expect that $ZT$ could see up to an order of magnitude improvement.

## 4. Conclusions

Using the fully quantum mechanical Non-Equilibrium Green's Function method, we calculated the electronic and thermoelectric coefficients of 2D channels embedded with nanoinclusions and voids. We found that while nanoinclusions and voids can have a positive impact on the Seebeck coefficient, the overall effect on the power factor is limited. We show that the power factor is resilient to variations in nanoinclusion/void density at all barrier heights in the ballistic regime, while under acoustic phonon scattering, the power factor is resilient to variable nanoinclusion/void density only at low barrier heights. We also show that the effect of voids on the power factor is dependent primarily on void density, and independent of void diameter.

**Acknowledgements**

This work has received funding from the European Research Council (ERC) under the European Union's Horizon 2020 Research and Innovation Programme (Grant Agreement No. 678763). M.T. has been supported by the Austrian Research Promotion Agency (FFG) Project No. 850743 QTSMoS.

**References**

[1] M. Zebarjadi, K. Esfarjani, M. S. Dresselhaus, Z. F. Ren and G. Chen, *Energy and Environmental Science,* 5 (2012) 5147-5162.
[2] K. Biswas, J. He, I. D. Blum, C.-I. Wu, T. P. Hogan, D. N. Seidman, V. P. Dravid and M. G. Kanatzidis, *Nature,* 489 (2012) 414-8.
[3] C. Gayner and K. K. Kar, *Progress in Materials Science,* 83, (2016) 330-382.
[4] T. Zou, X. Qin, Y. Zhang, X. Li, Z. Zeng, D. Li, J. Zhang, H. Xin, W. Xie and A. Weidenkaff, *Scientific Reports,* 5, (2015) 17803.
[5] P. E. Hopkins, C. M. Reinke, M. F. Su, R. H. Olsson, E. A. Shaner, Z. C. Leseman, J. R. Serrano, L. M. Phinney and I. El-Kady, *Nano Letters,* 11, (2011) 107-112.
[6] C. J. Vineis, A. Shakouri, A. Majumdar and M. G. Kanatzidis, *Advanced Materials,* 22, (2010) 3970-3890.
[7] A. Popescu, L. Woods, J. Martin and G. Nolas, *Physical Review B,* 79, (2009) p. 205302.
[8] A. Minnich and G. Chen, *Applied Physics Letters,* 91 (2007) 073105.
[9] M. Verdier, K. Termentzidis and D. Lacroix, *Journal of Applied Physics,* 119 (2016) 17.
[10] S. Wolf, N. Neophytou, Z. Stanojevic and H. Kosina, *Journal of Electronic Materials,* 43 (2014) 3870-3875.
[11] S. Wolf, N. Neophytou and H. Kosina, *Journal of Applied Physics,* 115 (2014) 204306.
[12] S. Fan, J. Zhao, Q. Yan, J. Ma and H. H. Hng, *Journal of Electronic Materials,* 40 (2011) 1018-1023.
[13] S. Ahmad, A. Singh, A. Bohra, R. Basu, S. Bhattacharya, R. Bhatt, K. N. Meshram, M. Roy, S. K. Sarkar, Y. Hayakawa, A. K. Debnath, D. K. Aswal and S. K. Gupta, *Nano Energy,* 27 (2016) 282-297.
[14] M. K. Keshavarz, D. Vasilevskiy, R. A. Masut and S. Turenne, *Materials Characterization,* 95 (2014) 44-49.
[15] M. Liu and X. Y. Qin, *Applied Physics Letters,* 101, (2012).
[16] T. Zou, X. Qin, Y. Zhang, X. Li, Z. Zeng, D. Li, J. Zhang, H. Xin, W. Xie and A. Weidenkaff, *Scientific Reports,* 5 (2015) 17803.
[17] T. H. Zou, X. Y. Qin, D. Li, G. L. Sun, Y. C. Dou, Q. Q. Wang, B. J. Ren, J. Zhang, H. X. Xin and Y. Y. Li, *Applied Physics Letters,* 104 (2014) 13904.
[18] J. Peng, L. Fu, Q. Liu, M. Liu, J. Yang, D. Hitchcock, M. Zhou and J. He, *J. Mater. Chem. A,* 2 (2014) 73-79.
[19] J. Zhou and R. Yang, *Journal of Applied Physics,* 110 (2011) 084317.
[20] S. Foster, M. Thesberg and N. Neophytou, *Physical Review B,* 96 (2017) 195425
[21] S. O. Koswatta, S. Hasan, M. S. Lundstrom, M. P. Anantram and D. E. Nikonov, *Ieee Transactions on Electron Devices,* 54 (2007) 2339-2351.
[22] S. Datta, Electronic Transport in Mesoscopic Systems, Cambridge MA: Cambridge University Press, 1997.
[23] M. P. Anantram, M. Lundstrom and D. E. Nikonov, *Proceedings of the IEEE,* 96 (2008) 1511-1550.
[24] M. P. L. Sancho, J. M. L. Sancho and J. Rubio, *Journal of Physics F: Metal Physics,* 15 (1985) 851-858.
[25] M. Thesberg, M. Pourfath, N. Neophytou and H. Kosina, *Journal of Electronic Materials,* 45, (2016) 1584-1588.
[26] N. Neophytou and H. Kosina, *Physical Review B*, 84 (2011) 085313.
[27] B. Qiu et al, *Europhysics Letters,* 109 (2015) 57006.
[28] M. P. Persson, Y.-m. Niquet, F. Triozon and S. Roche, *Nanoletters,* 8 (2008) 4146-4150.
[29] M. Dunham, B. Lorenzi, S. Andrews, A. Sood, M. Asheghi, D. Narducci, K. Goodson, *Applied Physics Letters*, 109 (2016) 253104
[30] J. Lee, J. Grossman, J. Reed, G. Galli, *Applied Physics Letters*, 91 (2007) 223110
[31] Y.He, D. Donadio, G. Galli, *Nano Letters,* 11 (2011) 3608-3611
[32] M. Samanta, S. Roychowdhury, J. Ghatak, P. Suresh, K. Biswas, *Chemistry – A European Journal*, 23 (2017) 7438-7443
[33] G. Zhu, H. Lee, Y. Lan, Y. C. Lan, D. Z. Wang, J. Yang, D. Vashaee, H. Guilbert, A. Pillitteri, M. S. Dresselhaus, G. Chen, Z. F. Ren, *Physical Review Letters*, 102 (2009) 196803